\documentclass[twocolumn,showpacs]{revtex4}

\usepackage{graphicx}
\usepackage{dcolumn}
\usepackage{amsmath}

\begin{document}

\markboth{}{
QUANTUM MONTE CARLO DIAGONALIZATION FOR ...~~~~~~~~~~~~~
PHYSICAL REVIEW B75, 224503 (2007)
}

\title{\small{PHYSICAL REVIEW B75, 224503 (2007)}\\
~~\\
\large{Quantum Monte Carlo diagonalization for many-fermion systems}
}

\author{Takashi Yanagisawa$^{a,b}$} 

\affiliation{$^a$Condensed-Matter Physics Group, Nanoelectronics Research 
Institute,
National Institute of Advanced Industrial Science and Technology (AIST),
Central 2, 1-1-1 Umezono, Tsukuba 305-8568, Japan\\
$^b$CREST, Japan Science and Technology Agency (JST),
Kawaguchi-shi, Saitama 332-0012, Japan
}

\date{2006; revised January 2007}

\begin{abstract}
In this study we present an optimization method based on the quantum Monte Carlo
diagonalization for many-fermion systems.
Using the Hubbard-Stratonovich transformation, employed to decompose the
interactions in terms of auxiliary fields, we expand
the true ground-state wave function.
The ground-state wave function is 
written as a linear combination of the basis wave functions. 
The Hamiltonian is diagonalized to obtain the lowest energy state, using
the variational principle within
the selected subspace of the basis functions.
This method is free from the difficulty known 
as {\em the negative sign problem}.  
We can optimize a wave function using two procedures.
The first procedure is to increase the number of basis functions. The second
improves each basis function through the operators,
$e^{-\Delta\tau H}$, using the Hubbard-Stratonovich decomposition.
We present an algorithm for the Quantum Monte Carlo diagonalization method 
using a genetic
algorithm and the renormalization method.
We compute the ground-state energy and correlation functions of small
clusters to compare with available data.

\end{abstract}


\maketitle

\section{Introduction}
The effect of the strong correlation between electrons is important for many 
quantum critical phenomena, such as unconventional superconductivity
(SC) and the metal-insulator transition.  
Typical correlated electron systems are high-temperature
superconductors\cite{dag94,sca90,and97,mor00}, 
heavy fermions\cite{ste84,lee86,ott87,map00} and organic 
conductors\cite{ish98}.
Recently the mechanisms of 
superconductivity in high-temperature superconductors and organic 
superconductors have been extensively studied using various two-dimensional 
(2D) models of electronic interactions.  
Among them the 2D Hubbard model\cite{hub63} is the simplest and most 
fundamental model.
This model has been studied intensively using numerical tools, such as the 
Quantum
Monte Carlo method
\cite{hir83,hir85,sor88,whi89,ima89,sor89,loh90,mor91,fur92,mor92,fah91,zha97,zha97b,kas01},
and the variational Monte Carlo 
method\cite{yok87,gro87,nak97,yam98,yan01,yan02,yan03,yan05,miy04}.
Recently, the two-leg ladder Hubbard model was also investigated with respect
to the mechanism of high-temperature 
superconductivity\cite{yam94,yam94b,koi99,noa96,noa97,kur96,dau00,san05}.

The Quantum Monte Carlo (QMC) method is a numerical method employed to 
simulate the behavior of correlated electron systems. 
It is well known, however, that there are significant issues associated with
the application to the QMC.
First, the standard Metropolis
(or heat bath) algorithm is associated with the negative sign problem.  
Second, the convergence of the trial wave function is sometimes not monotonic,
and further, is sometimes slow.
In past studies workers have investigated the possibility of eliminating
the negative sign 
problem\cite{fah91,zha97,kas01}.
If the negative sign problem can be eliminated, the next task would be to
improve the convergence of the simulation method.

In this paper we present an optimization method based on  Quantum Monte 
Carlo  diagonalization (QMD or QMCD).
The recent developments of high-performance computers have lead to the
possibility of the simulation of correlated electron systems using
diagonalization.
Typically, and as in this study, the ground-state wave function is defined as
\begin{equation}
\psi= e^{-\tau H}\psi_0,
\end{equation}
where $H$ is the Hamiltonian and $\psi_0$ is the initial one-particle
state such as the Fermi sea.
In the QMD method this wave function is written as a 
linear combination of the basis states, generated using the auxiliary 
field method based on the Hubbard-Stratonovich transformation; that is  
\begin{equation}
\psi= \sum_mc_m\phi_m,
\end{equation}
where $\phi_m$ are basis functions.
In this work we have assumed a subspace with $N_{states}$ basis wave functions.
From the variational principle, the coefficients $\{c_m\}$ are
determined from the diagonalization of the Hamiltonian, 
to obtain the lowest energy state in the selected subspace $\{\phi_m\}$.  
Once the $c_m$ coefficients are determined, the ground-state energy and 
other quantities
are calculated using this wave function.
If the expectation values are not highly sensitive to the number of basis
states, we can obtain the correct expectation values using an extrapolation
in terms of the basis states at the limit $N_{states}\rightarrow\infty$.
However, a more reliable procedure must be employed when the change in the 
values at the limit is not monotonic. 
In this study results are compared to results obtained from an exact 
diagonalization of small clusters, such as
4$\times$4 and $6\times 2$ lattices.

In the following section, Section II, we briefly review the
standard Quantum Monte Carlo simulation approach.
In Section III a discussion of the Quantum Monte Carlo diagonalization,
and an extrapolation method to obtain the expectation values, are presented.
Section IV is a discussion of the optimization procedure
which employs the diagonalization method.
All the results obtained in this study are compared to the exact and 
available results of small
systems in Section V.
Finally, a summary of the work presented in this paper is presented in 
Section VI.



\section{Quantum Monte Carlo Method}
The method of Quantum Monte Carlo diagonalization lies in the
QMC method.  Thus it is appropriate to first outline the QMC method.
The Hamiltonian is the Hubbard model containing on-site Coulomb 
repulsion and is written as

\begin{eqnarray}
H&=&-\sum_{ij\sigma}t_{ij}(c^{\dag}_{i\sigma}c_{j\sigma}+h.c.)
+ U\sum_jn_{j\uparrow}n_{j\downarrow},
\label{hamil}
\end{eqnarray}
where $c^{\dag}_{j\sigma}$ ($c_{j\sigma}$) is the creation 
(annihilation) operator of an electron with spin $\sigma$
at the $j$-th site and $n_{j\sigma}=c^{\dag}_{j\sigma}c_{j\sigma}$.  
$t_{ij}$ is the transfer energy between the sites $i$ and $j$.
$t_{ij}=t$ for the nearest-neighbor bonds. For all other cases $t_{ij}=0$.
$U$ is the on-site Coulomb energy. 
The number of sites is $N$ and
the linear dimension of the system is denoted as $L$.
The energy unit is given by $t$ and
the number of electrons is denoted as $N_e$.

In a Quantum Monte Carlo simulation, the ground state wave function is
\begin{equation}
\psi= {\rm e}^{-\tau H}\psi_0 ,
\label{psi}
\end{equation}
where $\psi_0$ is the initial one-particle state represented by a Slater 
determinant.  
For large $\tau$, ${\rm e}^{-\tau H}$ will project out the ground state from
$\psi_0$.
We write the Hamiltonian as $H=K+V$ where
K and V are the kinetic and interaction terms of the Hamiltonian in 
Eq.(\ref{hamil}), respectively. 
The wave function in Eq.(\ref{psi}) is written as
\begin{equation}
\psi= ({\rm e}^{-\Delta\tau (K+V)})^M \psi_0 \approx 
({\rm e}^{-\Delta\tau K}{\rm e}^{-\Delta\tau V})^M \psi_0 ,
\end{equation}
for $\tau=\Delta\tau\cdot M$.
Using the Hubbard-Stratonovich transformation\cite{hir83,bla81}, we have
\begin{eqnarray}
{\rm exp}(-\Delta\tau Un_{i\uparrow}n_{i\downarrow})
&=& \frac{1}{2}\sum_{s_i=\pm 1}{\rm exp}( 2as_i(n_{i\uparrow}
-n_{i\downarrow})\nonumber\\
&-&\frac{1}{2}U\Delta\tau(n_{i\uparrow}+n_{i\downarrow}) ),
\end{eqnarray}
for $({\rm tanh}a)^2={\rm tanh}(\Delta\tau U/4)$ or
${\rm cosh}(2a)={\rm e}^{\Delta\tau U/2}$.
The wave function is expressed as a summation of the one-particle Slater
determinants
over all the configurations of the auxiliary fields 
$s_j=\pm 1$.
The exponential operator is expressed as
\begin{eqnarray}
({\rm e}^{-\Delta\tau K}{\rm e}^{-\Delta\tau V})^M
&=& \frac{1}{2^{NM}}\sum_{\{s_i(\ell)\}}\prod_{\sigma}B_M^{\sigma}(s_i(M))
\nonumber\\
&\times& B_{M-1}^{\sigma}(s_i(M-1))\cdots B_1^{\sigma}(s_i(1)),\nonumber\\
\end{eqnarray}
where we have defined
\begin{equation}
B_{\ell}^{\sigma}(\{s_i(\ell)\})={\rm e}^{-\Delta\tau K_{\sigma}}
{\rm e}^{-V_{\sigma}(\{s_i(\ell)\})},
\label{bmat}
\end{equation}
for
\begin{equation}
V_{\sigma}(\{s_i\})= 2a\sigma\sum_i s_in_{i\sigma}-\frac{1}{2}
U\Delta\tau \sum_in_{i\sigma},
\end{equation}
\begin{equation}
K_{\sigma}=-\sum_{ij}t_{ij}(c_{i\sigma}^{\dag}c_{j\sigma}+h.c.).
\end{equation}
The ground-state wave function is
\begin{equation}
\psi= \sum_mc_m\phi_m ,
\label{wf}
\end{equation}
where $\phi_m$ is a Slater determinant corresponding to a configuration 
$m=\{s_i(\ell)\}$ ($i=1,\cdots,N; \ell=1,\cdots,M$)
of the auxiliary fields:
\begin{eqnarray}
\phi_m&=& \prod_{\sigma}B_M^{\sigma}(s_i(M))\cdots B_1^{\sigma}(s_i(1))\psi_0
\nonumber\\
&\equiv& \phi_m^{\uparrow}\phi_m^{\downarrow}.
\end{eqnarray}
The coefficients $c_m$ are constant real numbers: $c_1=c_2=\cdots$.
The initial state $\psi_0$ is a one-particle state.
If electrons occupy the wave numbers $k_1$, $k_2$, $\cdots$, $k_{N_{\sigma}}$
for each spin $\sigma$, $\psi_0$ is given by the product 
$\psi_0^{\uparrow}\psi_0^{\downarrow}$ where $\psi_0^{\sigma}$ is the matrix
represented as\cite{ima89}
\begin{equation}
\left( \begin{array}{ccccc}
{\rm e}^{ik_1\cdot r_1} & {\rm e}^{ik_2\cdot r_1} & \cdots & \cdots &
{\rm e}^{ik_{N_{\sigma}}\cdot r_1}  \\
{\rm e}^{ik_1\cdot r_2} & {\rm e}^{ik_2\cdot r_2} & \cdots & \cdots &\cdots\\  
\cdot & \cdot & \cdot & \cdot & \cdot \\
{\rm e}^{ik_1\cdot r_N} & {\rm e}^{ik_2\cdot r_N} & \cdots & \cdots & 
\end{array} \right).
\end{equation}
$N_{\sigma}$ is the number of electrons for spin $\sigma$.
In actual calculations we can use a real representation where the matrix 
elements are cos$(k_i\cdot r_j)$ or sin$(k_i\cdot r_j)$.
In the real-space representation, the matrix of $V_{\sigma}(\{s_i\})$ is a
diagonal matrix given as
\begin{equation}
V_{\sigma}(\{s_i\})={\rm diag}(2a\sigma s_1-U\Delta\tau/2,\cdots,
2a\sigma s_N-U\Delta\tau/2).
\end{equation}
The matrix elements of $K_{\sigma}$ are
\begin{eqnarray}
(K_{\sigma})_{ij}&=& -t~~~i,j~ {\rm are~ nearest~ neighbors}\nonumber\\
&=& 0~~~{\rm otherwise}.
\end{eqnarray}
$\phi_m^{\sigma}$ is an $N\times N_{\sigma}$ matrix given by the product
of the matrices ${\rm e}^{-\Delta\tau K_{\sigma}}$, ${\rm e}^{V_{\sigma}}$
and $\psi_0^{\sigma}$.
The inner product is thereby calculated as a determinant\cite{zha97},
\begin{equation}
\langle\phi_m^{\sigma}\phi_n^{\sigma}\rangle=
{\rm det}(\phi_m^{\sigma\dag}\phi_n^{\sigma}).
\end{equation}
The expectation value of the quantity $Q$ is evaluated as
\begin{equation}
\langle Q\rangle = \frac{\sum_{mn}\langle\phi_m Q\phi_n\rangle}
{\sum_{mn}\langle\phi_m\phi_n\rangle}.
\label{qexpe}
\end{equation}
If $Q$ is a bilinear operator $Q_{\sigma}$ for spin $\sigma$, we have
\begin{eqnarray}
\langle Q_{\sigma}\rangle &=& \frac{
\sum_{mn}\langle\phi_m^{\sigma}Q_{\sigma}\phi_n^{\sigma}\rangle
\langle\phi_m^{-\sigma}\phi_n^{-\sigma}\rangle}
{\sum_{mn}\langle\phi_m^{\sigma}\phi_n^{\sigma}\rangle
\langle\phi_m^{-\sigma}\phi_n^{-\sigma}\rangle}
\nonumber\\
&=& \frac{\sum_{mn}\langle\phi_m^{\sigma}Q_{\sigma}\phi_n^{\sigma}\rangle
{\rm det}(\phi_m^{-\sigma\dag}\phi_n^{-\sigma})}
{\sum_{mn}{\rm det}(\phi_m^{\sigma\dag}\phi_n^{\sigma})
{\rm det}(\phi_m^{-\sigma\dag}\phi_n^{-\sigma})}
\nonumber\\
&=& \sum_{mn}\frac{{\rm det}(\phi_m^{\sigma\dag}\phi_n^{\sigma})
{\rm det}(\phi_m^{-\sigma\dag}\phi_n^{-\sigma})}
{\sum_{m'n'}{\rm det}(\phi_{m'}^{\sigma\dag}\phi_{'n}^{\sigma})
{\rm det}(\phi_{m'}^{-\sigma\dag}\phi_{n'}^{-\sigma})}\nonumber\\
&\times& \frac{\langle\phi_m^{\sigma}Q_{\sigma}\phi_n^{\sigma}\rangle}
{\langle\phi_m^{\sigma}\phi_n^{\sigma}\rangle}.
\end{eqnarray}
The expectation value with respect to the Slater determinants
$\langle\phi_m^{\sigma}Q_{\sigma}\phi_n^{\sigma}\rangle$ is evaluated
using the single-particle Green's function\cite{ima89,zha97},
\begin{equation}
\frac{\langle\phi_m^{\sigma}c_{i\sigma}c_{j\sigma}^{\dag}\phi_n^{\sigma}
\rangle}
{\langle\phi_m^{\sigma}\phi_n^{\sigma}\rangle}
=\delta_{ij}-(\phi_n^{\sigma}
(\phi_m^{\sigma\dag}\phi_n^{\sigma})^{-1}\phi_m^{\sigma\dag})_{ij}.
\end{equation}
In the above expression, 
$P_{mn}\equiv {\rm det}(\phi_m^{\sigma}\phi_n^{\sigma})
{\rm det}(\phi_m^{-\sigma}\phi_n^{-\sigma})$
can be regarded as the weighting factor to obtain the Monte Carlo samples.
Since this quantity is not necessarily positive definite, the weighting factor
should be $|P_{mn}|$; the resulting relationship is,
\begin{eqnarray}
\langle Q_{\sigma}\rangle&=& \sum_{mn}P_{mn}\langle Q_{\sigma}\rangle_{mn}
/\sum_{mn}P_{mn}\nonumber\\
&=& \sum_{mn}|P_{mn}|sign(P_{mn})\langle Q_{\sigma}\rangle_{mn}
/\sum_{mn}|P_{mn}|sign(P_{mn})\nonumber\\
\end{eqnarray}
where
$sign(a)=a/|a|$ and
\begin{equation}
\langle Q_{\sigma}\rangle_{mn}=\frac{\langle\phi_m^{\sigma}Q_{\sigma}
\phi_n^{\sigma}\rangle}{\langle\phi_m^{\sigma}\phi_n^{\sigma}\rangle}.
\end{equation} 
This relation can be evaluated using a Monte Carlo procedure if an 
appropriate algorithm, such as the Metropolis or heat bath method, is 
employed\cite{bla81}.
The summation can be evaluated using appropriately defined Monte Carlo
samples,
\begin{equation}
\langle Q_{\sigma}\rangle= \frac{ \frac{1}{n_{MC}}\sum_{mn}sign(P_{mn})
\langle Q_{\sigma}\rangle_{mn}}{\frac{1}{n_{MC}}\sum_{mn}sign(P_{mn})},
\label{qqmc}
\end{equation}
where $n_{MC}$ is the number of samples.
The sign problem is an issue if the summation of $sign(P_{mn})$
vanishes within statistical errors.  In this case it is indeed impossible
to obtain definite expectation values.

\section{Quantum Monte Carlo Diagonalization}

\subsection{Diagonalization}
Quantum Monte Carlo diagonalization (QMD) is a method for the evaluation of
$\langle Q_{\sigma}\rangle$ without {\em the negative sign problem}.
The configuration space of the probability $\|P_{mn}\|$ in Eq.(\ref{qqmc}) 
is generally very strongly peaked.
The sign problem lies in the distribution of $P_{mn}$ in the configuration
space.
It is important to note that the distribution of the basis functions $\phi_m$
($m=1,2,\cdots$) is uniform since $c_m$ are constant numbers: $c_1=c_2=\cdots$.
In the subspace $\{\phi_m\}$, selected from all configurations of auxiliary
fields, the right-hand side of Eq.(\ref{qexpe}) can be determined.
However, the large number of basis states required to obtain accurate
expectation values is beyond the current storage capacity of computers.
Thus we use the variational principle to obtain the expectation values.

From the variational principle,
\begin{equation}
\langle Q\rangle = \frac{\sum_{mn}c_mc_n\langle\phi_m Q\phi_n\rangle}
{\sum_{mn}c_mc_n\langle\phi_m\phi_n\rangle},
\end{equation}
where $c_m$ ($m=1,2,\cdots$) are variational parameters.
In order to minimize the energy
\begin{equation}
E = \frac{\sum_{mn}c_mc_n\langle\phi_m H\phi_n\rangle}
{\sum_{mn}c_mc_n\langle\phi_m\phi_n\rangle},
\end{equation}
the equation $\partial E/\partial c_n=0$ ($n=1,2,\cdots$) is solved for,
\begin{equation}
\sum_m c_m\langle\phi_n H\phi_m\rangle-E\sum_m c_m\langle\phi_n\phi_m\rangle
=0.
\end{equation}
If we set
\begin{equation}
H_{mn}=\langle\phi_m H\phi_n\rangle,
\end{equation}
\begin{equation}
A_{mn}=\langle\phi_m\phi_n\rangle,
\end{equation}
the eigen equation is
\begin{equation}
Hu=EAu,
\end{equation}
for $u=(c_1,c_2,\cdots)^t$.
Since $\phi_m$ ($m=1,2,\cdots$) are not necessarily orthogonal,
$A$ is not a diagonal matrix.
We diagonalize the Hamiltonian $A^{-1}H$, and then
calculate the expectation values of correlation functions 
with the ground state eigenvector;  
in general $A^{-1}H$ is not a symmetric matrix.

In order to optimize the wave function we must increase the number of
basis states $\{\phi_m\}$.
This can be simply accomplished through random sampling.
For systems of small sizes and small $U$, we can evaluate the expectation
values from an extrapolation of the basis of randomly generated states.

\subsection{Extrapolation}
  In Quantum Monte Carlo simulations an extrapolation is performed to obtain 
the expectation values for the ground-state wave function.  
If $M$ is large enough, the wave function in Eq.(\ref{wf}) will approach
the exact ground-state wave function, $\psi_{exact}$, as the number of basis 
functions, $N_{states}$, is increased.
If the number of basis functions is large enough, the wave function
will approach, $\psi_{exact}$, as $M$ is increased.
In either case the method  employed for the reliable 
extrapolation of the wave function is a key issue in calculating the
expectation values.  
If the convergence is fast enough, the expectation values can be obtained
from the extrapolation in terms of $1/N_{states}$.
Note that although the extrapolation in 
terms of 1/$M$, or the time step $\Delta\tau$, has often been employed in QMC 
calculations, however, 
a linear dependence for 1/$M$ or $\Delta\tau$ will not 
necessarily guarantee. an accurate extrapolated result.  
The variance method was recently proposed in
variational and Quantum Monte Carlo simulations, where the extrapolation is
performed as a function of the variance.
An advantage of the variance method lies is that
linearity is expected in some
cases\cite{sor01,kas01}:
\begin{equation}
\langle Q\rangle-Q_{exact}\propto v,
\end{equation}
where $v$ denotes the variance defined as
\begin{equation}
v= \frac{\langle (H-\langle H\rangle)^2\rangle}{\langle H\rangle^2}
\label{vari}
\end{equation}
and $Q_{exact}$ is the expected exact value of the quantity $Q$.

The following brief proof clearly shows that the energy in Eq.(\ref{vari})
varies linearly.
If we denote the exact ground-state wave function as $\psi_g$
and the excited states as $\psi_i$ ($i=1,2,\cdots$),  the wave function can be
written as
\begin{equation}
\psi= a\psi_g+\sum_ib_i\psi_i,
\end{equation}
where we assume that $a$ and $b_i$ are real and satisfy $a^2+\sum_ib_i^2=1$.
If it is assumed that $H\psi_g=E_g\psi_g$ and $H\psi_i=E_i\psi_i$, 
the energy is found to be
\begin{eqnarray}
E&=& \langle H\rangle\nonumber\\
&=& a^2\langle\psi_g H\psi_g\rangle+2a\sum_ib_i\langle\psi_i H\psi_g\rangle
+\sum_{ij}b_ib_j\langle\psi_i H\psi_j\rangle\nonumber\\
&=& a^2E_g+\sum_{ij}b_ib_j\langle\psi_iH\psi_j\rangle\nonumber\\
&=& a^2E_g+\sum_ib_i^2E_i.
\end{eqnarray}
The deviation of $E$ from $E_g$ is
\begin{eqnarray}
\delta E&=& E-E_g\nonumber\\
&=& (a^2-1)E_g+\sum_ib_i^2E_i\nonumber\\
&=& b^2(\langle E_i\rangle-E_g)
\end{eqnarray}
where $b^2=1-a^2$ and $\langle E_i\rangle=\sum_jb_j^2E_j/\sum_jb_j^2$.
The variance $v$ of $H$ is also shown to be proportional to $b^2$ if
$b^2$ is small.  
Since $\langle H^2\rangle=a^2E_g+b^2\langle E_i^2\rangle$ where
$\langle E_i^2\rangle=\sum_jb_j^2E_j^2/\sum_jb_j^2$, $v$ is evaluated as
\begin{equation}
v= C\{ (1-b^2)\frac{\delta E}{E_g}-2\left(\frac{\delta E}{E_g}\right)^2
+\cdots\},
\end{equation}
for a constant $C$.  Hence if $b$ is small it is found that
\begin{equation}
\frac{\delta E}{E_g}= \frac{v}{C}+O(v^2).
\end{equation}
The other quantities can be found if
$Q_g=\langle\psi_g Q\psi_g\rangle$, which leads to the result
\begin{equation}
\langle Q\rangle-Q_g= -b^2Q_g+2a\sum_ib_i\langle\psi_i Q\psi_g\rangle
+\sum_{ij}b_ib_j\langle\psi_i Q\psi_j\rangle.
\end{equation}
If $Q$ commutes with $H$, and $\psi_i$ are eigenstates of $Q$, 
$\langle Q\rangle-Q_g$ is proportional to $b^2$.
\begin{equation}
\langle Q\rangle-Q_g= -b^2(Q_g-\langle Q_i\rangle),
\end{equation}
where $\langle Q_i\rangle=\sum_ib_i^2\langle\psi_i Q\psi_i\rangle/\sum_ib_i^2$;
thus $\langle Q\rangle-Q_g\propto v$.
In the general case $[H,Q]\neq 0$, $\langle Q\rangle-Q_g$ is not
necessarily proportional to $b^2$.  However, if the matrix element 
$\langle\psi_i Q\psi_g\rangle$ is negligible, we obtain
\begin{eqnarray}
\langle Q\rangle-Q_g&=& -b^2Q_g+\sum_{ij}b_ib_j\langle\psi_i Q\psi_j\rangle
\nonumber\\
&=& -b^2(Q_g-\sum_{ij}b_ib_j\langle\psi_i Q\psi_j\rangle/\sum_ib_i^2).
\end{eqnarray}
This shows that $\langle Q\rangle-Q_g$ is proportional to the variance $v$.
Thus, if $\langle\psi_i Q\psi_g\rangle$ is small, we can perform an 
extrapolation
using a linear fit to obtain the expectation values.
We expect that this is the case for short-range correlation functions,
since the local correlation may give rise to small effects in the
orthogonality of $\psi_i$ and $\psi_g$, i.e. $\langle\psi_i\psi_g\rangle=0$.
Hence the evaluations of local quantities will be much easier than for the
long-range correlation functions.

\section{Optimization in Quantum Monte Carlo Diagonalization}

\subsection{Simplest algorithm}
The simplest procedure for optimizing the ground-state wave function is to
increase the number of basis states $\{\phi_m\}$ by random sampling.
First, we set $\tau$ and $M$, for example, $\tau=0.1$, 0.2, $\cdots$, and
$M=20$, 30, $\cdots$.
We denote the number of basis functions as $N_{states}$.  
We start with $N_{states}=100\sim 300$ and then increase up to 2000 or 3000.
This procedure can be outlined as follows:\\
\\
A1. Generate the auxiliary fields $s_i$ ($i=1,\cdots,N$) in 
$B_{\ell}^{\sigma}(\{s_i\}))$  randomly for $\ell=1,\cdots,M$ for
$\phi_m$ ($m=1,\cdots,N_{states}$), and
generate $N_{states}$ basis wave function $\{\phi_m\}$.\\
A2. Evaluate the matrices $H_{mn}=\langle\phi_m H\phi_n\rangle$ and
$A_{mn}=\langle\phi_m\phi_n\rangle$, and diagonalize the matrix
$A^{-1}H$ to obtain $\psi=\sum_m c_m\phi_m$.  
Then calculate the expectation values and the energy
variance.\\
A3. Repeat the procedure from A1 after increasing the number of basis 
functions.\\
\\
For small systems this random method produces reliable energy  results.
The diagonalization plays an importance producing fast convergence.

Failure of this simple method sometimes occurs as the system size is increased.
The eigenfunction of $A^{-1}H$ can be localized when the off-diagonal elements
are small,  meaning that
some components of $c_m$ are large and others are negligible.
A quotient of localization in the configuration space can be defined.
For example, the summation of $\left|c_m\right|^2$ except $\phi_n$ with large 
$c_n$
is a candidate for such property,
\begin{equation}
Q_{loc}= \sum_m'\left|c_m\right|^2,
\label{qloc}
\end{equation}
where the prime indicates that the summation is performed excluding the 
largest $c_n$. 
$Q_{loc}$ should approach 1 as the number of basis functions is
increased. 
In the case of localization, $Q_{loc}< 0.1$, where to lower the energy is
procedurally inefficient.
In order to avoid the localization difficulty there are two possible procedures.
First is to multiply $\phi_m$ by $B_{\ell}^{\sigma}(\{s_i\}))$ to improve 
and optimize the basis 
wave function $\phi_m$ further.  Second, use a more effective method 
to generate
new basis functions, explained further in the subsequent sections.

\subsection{Renormalization}

The basis functions $\{\phi_m\}$ multiplied by $B_{\ell}^{\sigma}$
($\ell=M+1,M+2,\cdots$) are improved to provide a lower ground state.
Here the 'improvement' means the increase of $\tau$ in Eq.(\ref{psi}) which
is accomplished by increasing $M$.
The matrix $B_{\ell}^{\sigma}(\{s_i\}))$ is given by a summation over $2^N$
configurations of $\{s_i\}$.  If we consider all of these configurations,
the space required for basis functions becomes large.  
Thus, we should select several configurations or one configuration that
exhibits the lowest energy.
One procedure to choose such a state is the following:
\\
\\
R1. Multiply $\phi_m$ by
$\prod_{\sigma}{\rm exp}(2a\sigma s_jn_{j\sigma}-\frac{1}{2}U\Delta\tau
n_{j\sigma})$, where we generate the auxiliary fields $s_i(\ell)$ for $\ell=M+1$
and $i=1,\cdots,N$ using random numbers.
Then evaluate the ground state energy.  If the energy is lower, 
$\phi_m$ is defined as a new and improved basis function.
If we have a higher energy, $\phi_m$ remains unchanged.
Repeat this procedure to lower the ground state energy twenty to
fifty times.\\
R2. Repeat above for $m=1,\cdots,N_{states}$.\\
R3. Multiply $\phi_m$ by the kinetic operator $e^{-\Delta\tau K_{\uparrow}}$
and $e^{-\Delta\tau K_{\downarrow}}$.\\
R4. Repeat from R1 and continue for $\ell\rightarrow\ell+1$.\\
\\
This method is referred to as the $1/2^N$-method in this paper since one 
configuration is chosen from $2^N$ possible states.
It is important to note that $N_{states}$ remains unchanged.
An alternative method has been proposed to renormalize $\{\phi_m\}$ and
is outlined as\cite{kas01}:\\
\\
R'1. Multiply $\phi_m$ by $\prod_{\sigma}{\rm exp}(2a\sigma s_jn_{j\sigma}-\frac{1}{2}
U\Delta\tau n_{j\sigma})$ and evaluate the energy for
$s_j=1$ and $s_j=-1$.  We adopt $s_j$ for which we have the lower energy.\\
R'2. Repeat this procedure for $j=1,\cdots,N$ and determine the
configuration $\{s_j\}$ for $\phi_m$.\\
R'3. Multiply $\phi_m$ by the kinetic operator $e^{-\Delta\tau K_{\uparrow}}$
and $e^{-\Delta\tau K_{\downarrow}}$.\\
R'4. Repeat above for $m=1,\cdots,N_{states}$ to improve $\phi_m$, and
repeat from R1.\\
\\
In this latter method the energy is calculated for the auxiliary field $s_i=\pm 1$ 
at each site before making a selection.
In the literature\cite{kas01} this procedure is called the path-integral
renormalization group (PIRG) method.

\subsection{Genetic algorithm}

In order to lower the ground-state energy 
efficiently, we can employ a genetic algorithm\cite{gol89}  to generate
the basis set from the initial basis set.
One idea is to replace some parts of $\{s_i(\ell)\}$ 
($i=1,\cdots,N; \ell=1,\cdots,M$) in $\phi_n$ that has the large weight
$\left|c_n\right|^2$ to generate a new basis function $\phi'_n$.
The new basis function $\phi'_n$ obtained in this way is expected to also
have a
large weight and contribute to $\psi$.

Let us consider two basis functions $\phi_m$ and $\phi_n$
chosen from the basis set with a probability proportional to the weight 
$\left|c_j\right|^2$ using uniform random numbers.
For example, since $\sum_{all j}\left|c_j\right|^2=1$, we set the weight of
$\phi_{\ell}$ to occupy 
$\sum_{j=1}^{\ell-1}\left|c_j\right|^2<x<\sum_{j=1}^{\ell}\left|c_j\right|^2$
in the range $0<x<1$.  If the random number $r$ is within
$\sum_{j=1}^{m-1}\left|c_j\right|^2<r<\sum_{j=1}^m\left|c_j\right|^2$,
we choose $\phi_m$, and  $\phi_n$ is similarly chosen.
A certain part of the genetic data between $\phi_m$ and $\phi_n$ is exchanged, 
which results in two new basis functions $\phi'_m$ and $\phi'_n$.
We add $\phi'_n$, or $\phi'_m$, or both of them, to the set of basis functions
as new elements.
In this process every site is labeled using integers such as $i=1,\cdots,N$, 
and then we exchange $s_i$ for $i=L_1,L_1+1,\cdots,L_1+L_{exch}-1$ where
the number of $s_i$ to be exchanged is denoted as $L_{exch}$.
$L_1$ can be determined using random numbers.
We must also include a randomly generated new basis function as a
mutation.
Here we fix the numbers $N_{states}$ and $N_{step}$ before starting the Monte
Carlo steps.
For instance, $N_{states}=200$ and $N_{step}=200$.
$N_{states}$ is increased as the Monte Carlo steps progress.
We diagonalize the Hamiltonian $A^{-1}H$ at each step when the $N_{step}$ basis 
functions are added to the basis set
in order to recalculate the weight $\left|c_k\right|^2$
($k=1,2,\cdots$). 
The procedure is summarized as follows:
\\
\\
G1. Generate the auxiliary fields $s_i(\ell)$ ($i=1,\cdots,N$) randomly
for $\ell=1,\cdots.M$. 
Generate $N_{states}$ basis functions $\{\phi_k\}$.
This is the same as A1.\\
G2. Evaluate the matrices $H_{mn}=\langle\phi_m H\phi_n\rangle$ and
$A_{mn}=\langle\phi_m\phi_n\rangle$, and diagonalize the matrix
$A^{-1}H$ to obtain $\psi=\sum_m c_m\phi_m$ and 
calculate the expectation values and the energy variance.
This is the same as A2.\\
G3. Determine whether a new basis function should be generated randomly or 
using the genetic method on the basis of random numbers.  
Let $r_c$ be in the range $0<r_c<1$,  for example, $r_c=0.9$.
If the random number $r$ is less than $r_c$, a new basis function is defined 
using the genetic algorithm and the next step G4 is executed,
otherwise generate the auxiliary fields $\{s_i\}$ randomly and go to G6. \\
G4. The weight of $\phi_k$ is given as $\left|c_k\right|^2$.
Choose two basis functions $\phi_m$ and $\phi_n$ from the basis set
with a probability proportional to the weight $\left|c_k\right|^2$.
Now we determine which part of the genetic code is exchanged between
$\phi_m$ and $\phi_n$.
We choose $\ell=\ell_0$ for $1\leq\ell\leq M$ using random numbers. 
We choose the sites $j=L_1,\cdots,L_2=L_1+L_{exch}-1$ 
for a randomly chosen $L_1$.\\
G5. Exchange the genetic code $\{s_i(\ell)\}$ between $\phi_m$ and
$\phi_n$ for $\ell=\ell_0$ and $j=L_1,\cdots,L_2+L_{exch}-1$.  
We have two new functions $\phi'_m$ and $\phi'_n$.
We adopt one or two of them as basis functions and keep the originals
$\phi_m$ and $\phi_n$ in the basis set.\\
G6. If the $N_{step}$ basis functions are added up to the basis set after
step G2, then repeat from step G2,
otherwise repeat from step G3.
\\
\\

\begin{figure}
\includegraphics[width=10.5cm]{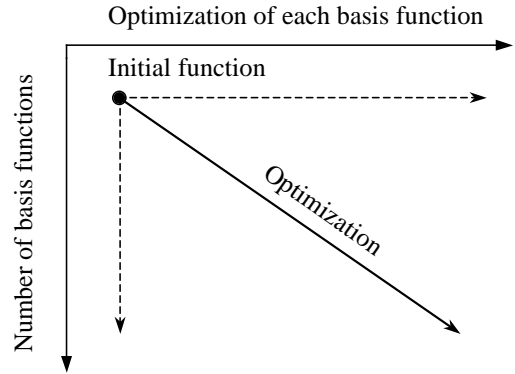}
\caption{
Concept of optimization procedure.
There are three approaches to reach the ground-state wave function.
First is to increase the number of basis functions for fixed $m$.
Second is to increase $M$ multiplying each basis function by 
$B_{\ell}(\{s_i\})$.  Third is the hybrid method of the previous two
procedures.
}
\label{opti}
\end{figure}

\subsection{Hybrid optimization algorithm}
In actual calculations it is sometimes better to use a hybrid of genetic 
algorithm
and renormalization method.
The concept to reach the ground-state wave function employed in
this study is presented in Fig.1.
There are two possible paths; one is to increase the number of basis 
functions using
the genetic algorithm and the other is to improve each basis function by
the matrix $B_{\ell}(\{s_i\})$.
The path followed when the hybrid procedure is employed is the average of
these two paths and is represented as the diagonal illustrated in Fig.1.
Before step G6 in the genetic algorithm, the basis functions $\phi_m$
are multiplied by $B_{\ell}(\{s_i\})$ following the renormalization algorithm
of the steps R1 to R3.  Then we go to G6.
The method is summarized as follows:\\
\\
H1. Generate the auxiliary fields $s_i(\ell)$ ($i=1,\cdots,N$) randomly
for $\ell=1,\cdots.M$.
Generate $N_{states}$ basis functions $\{\phi_k\}$.
\\
H2. Evaluate the matrices $H_{mn}=\langle\phi_m H\phi_n\rangle$ and
$A_{mn}=\langle\phi_m\phi_n\rangle$, and diagonalize the matrix
$A^{-1}H$ to obtain $\psi=\sum_m c_m\phi_m$ and
calculate the expectation values and the energy variance.
\\
H3. Determine whether a new basis should be generated randomly or
using the genetic algorithm.
Let $r_c$ be in the range  $0<r_c<1$.
If the random number $r$ is less than $r_c$, a new basis function is defined
using the genetic algorithm and the next step is H4,
otherwise generate the auxiliary fields $\{s_i\}$ randomly and go to H6. \\
H4. The weight of $\phi_k$ is given as $\left|c_k\right|^2$.
Choose two basis functions $\phi_m$ and $\phi_n$ from the basis set
with a probability proportional to the weight $\left|c_k\right|^2$.
Now we determine which part of the genetic code is exchanged between
$\phi_m$ and $\phi_n$.
We choose $\ell=\ell_0$ for $1\leq\ell\leq M$ using random numbers.
We choose the sites $j=L_1,\cdots,L_2=L_1+L_{exch}-1$
for a randomly chosen $L_1$.\\
H5. Exchange the genetic code $\{s_i\}$ between $\phi_m$ and
$\phi_n$ for $\ell=\ell_0$ and $j$ determined in step H4.  
We have two new functions $\phi'_m$ and $\phi'_n$.
We adopt one or two of them as basis functions and keep the originals
$\phi_m$ and $\phi_n$ in the basis set.\\
H6. Multiply $\phi_m$ by
$\prod_{\sigma}{\rm exp}(2a\sigma s_jn_{j\sigma}-\frac{1}{2}U\Delta\tau
n_{j\sigma})$, where we generate the auxiliary fields $s_i(\ell)$ for $\ell=M+1$and $i=1,\cdots,N$ using random numbers.
Then evaluate the ground state energy.  If the energy is lower,
$\phi_m$ is defined as a new and improved basis function.
If we have a higher energy, $\phi_m$ remains unchanged.
Repeat this procedure to lower the ground state energy twenty to
fifty times.\\
H7. Repeat above for $m=1,\cdots,N_{states}$.\\
H8. Multiply $\phi_m$ by the kinetic operator $e^{-\Delta\tau K_{\uparrow}}$
and $e^{-\Delta\tau K_{\downarrow}}$.\\
H9. If the $N_{step}$ basis functions are added up to the basis set after
step H2, then repeat from H2,
otherwise repeat from step H3.

\subsection{Discussion on the Quantum Monte Carlo Diagonalization}

The purpose of the QMD method is to calculate
\begin{equation}
\langle Q\rangle = \frac{\sum_{mn}c_mc_n\langle\phi_m Q\phi_n\rangle}
{\sum_{mn}c_mc_n\langle\phi_m\phi_n\rangle}.
\end{equation}
In an algorithm based on the Quantum Monte Carlo procedures, we evaluate the
expectation values in the subspace $\{\phi_i\}$, selected from all the
configurations of the auxiliary fields.
From the data showing how the mean values $\langle Q\rangle$
varies as the subspace is enlarged, we can estimate the exact value of
$\langle Q\rangle$ using an extrapolation.
A devised algorithm may help us to perform the Quantum
Monte Carlo evaluations efficiently.
We have presented the genetic algorithm and the renormalization method.
It may be possible to overcome the problem of
localization in the subspace using this algorithm.  In fact, the
quotient $Q_{loc}$ in Eq.(\ref{qloc}) becomes nearly 1, i.e. $Q_{loc}>0.99$,
in the evaluations presented in the next section.
For such a case, most of basis functions in the subspace give contributions 
to the mean values of physical quantities and the obtained results are
certainly reliable.


\begin{figure}
\includegraphics[width=\columnwidth]{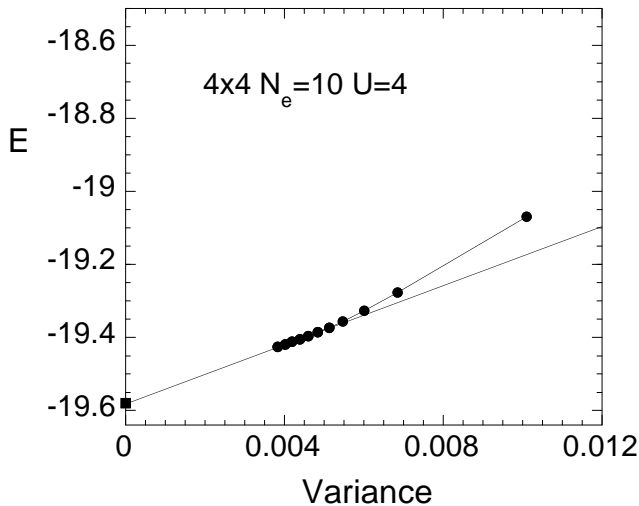}
\caption{
Energy as a function of the variance for $4\times 4$, $U=4$ and $N_e=10$.
The square is the exact result.
The data fit using a straight line using the least-square method
as the variance is reduced.
We started with $N_{states}=100$ (first solid circle) and then increase 
up to 2000.
}
\label{4x4-10}
\end{figure}

\begin{figure}
\includegraphics[width=\columnwidth]{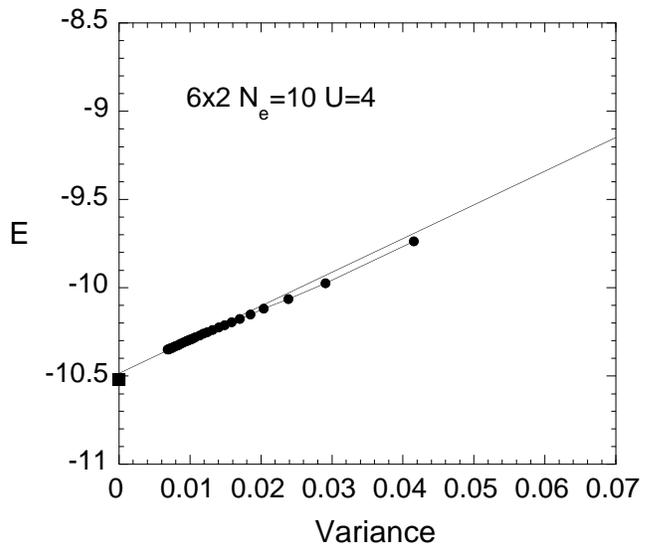}
\caption{
Energy as a function of the variance  for $6\times 2$
$N_e=10$ and $U=4$.  
The square is the exact value obtained using exact diagonalization.
}
\label{6x2-10}
\end{figure}

\begin{figure}
\includegraphics[width=10.5cm]{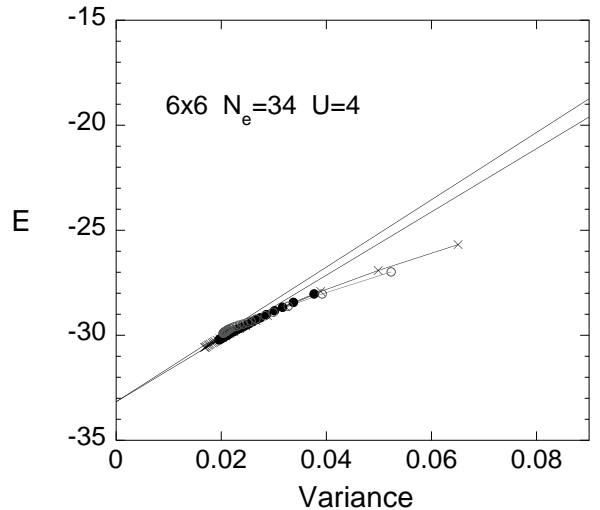}
\caption{
Energy as a function of the variance $v$ for $6\times 6$. 
with the periodic boundary conditions.
Solid circles and crosses are data obtained from the QMD method for
two different initial configurations of the auxiliary fields.
Gray open circles show results obtained from the $1/2^N$-renormalization
method (PIRG) with 300 basis wave functions. 
}
\label{6x6-34}
\end{figure}

\begin{figure}
\includegraphics[width=\columnwidth]{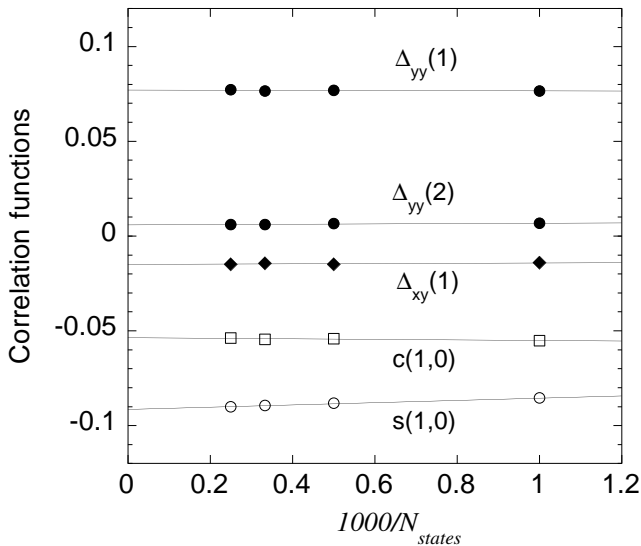}
\caption{
Correlation functions obtained by QMD for $4\times 4$ lattice with
$N_e=10$ and $U=4$ as a function of $1/N_{states}$.
}
\label{4x4-10C}
\end{figure}

\section{Results}

In this section, the results obtained using the QMD method are compared
to the exact and available results.
We investigate the small clusters (such as $4\times 4$ and $6\times 6$),
the one-dimensional (1D) Hubbard model, the ladder Hubbard model, and
the two-dimensional (2D) Hubbard model.

\subsection{Ground-state energy and correlation functions: 
check of the method}

The results for the $4\times 4$, $6\times 2$ and $6\times 6$ systems 
are presented in Table I.
The results are compared to the exact values and those available values
obtained
using the exact diagonalization, the quantum Monte Carlo method, 
the constrained path Monte Carlo method\cite{zha97} and the variational
Monte Carlo method for 
lattices with periodic boundary conditions.  
The expectation
values for the ground state energy are presented for several values of $U$.
The data include the cases for open shell structures where the highest-occupied
energy levels are partially occupied by electrons.
In the open shell cases the evaluations are sometimes extremely difficult. 
As is apparent from Table I, our method gives results in reasonable
agreement with the exact values.
The energy as a function of the variance is presented in Figs.2, 3 and 4.
To obtain these results the genetic algorithm was employed to produce the
basis functions except the open symbols in Fig.4.
The $4\times 4$ where $N_e=10$ in Fig.2 is the energy for the closed shell
case up to 2000 basis states.
The other two figures are for open shell cases, where evaluations were
performed up to 3000 states.
Open symbols in Fig.4 indicate the energy obtained using the renormalization
method ($1/2^N$-method) with 300 basis states.  The results for the QMD
and $1/2^N$-method (or PIRG) are quite similar as a function of the energy 
variance.
In these cases $Q_{loc}$ is close to $1$; $Q_{loc}\sim 0.99$.
As the variance is reduced, the data can fit using a straight line
using the least-square method.

In Table I we have also included the VMC results for the $\lambda$-functions.
The $\lambda$-functions are variational functions defined as follows.
The Gutzwiller function is well known as
\begin{equation}
\psi_G= P_G\psi_0,
\end{equation}
where $P_G$ is the Gutzwiller projection operator,
\begin{equation}
P_G= \prod_j[1-(1-g)n_{j\uparrow}n_{j\downarrow}].
\end{equation}
$g$ is the parameter in the range  $0\leq g \leq 1$.
The non-interacting wave function $\psi_0$ is optimized by controlling
the double occupancy $\sum_j\langle n_{j\uparrow}n_{j\downarrow}\rangle$.
The further optimization of the Gutzwiller function can be 
obtained\cite{oht92,yan98},
\begin{equation}
\psi_{\lambda}^{(1)}={\rm e}^{-\lambda K}{\rm e}^{-\alpha V}\psi_G,
\end{equation}
\begin{equation}
\psi_{\lambda}^{(2)}={\rm e}^{-\lambda' K}{\rm e}^{-\alpha' V}
\psi_{\lambda}^{(1)},
\end{equation}
where $K$ is the kinetic energy term and $V$ is the on-site
Coulomb interaction,
\begin{equation}
V= \sum_jn_{j\uparrow}n_{j\downarrow},
\end{equation}
where $\lambda$, $\alpha$, $\lambda'$, $\alpha'$ are variational
parameters to be determined, to lower the ground-state energy.
$\alpha$ is related to $g$ as $\alpha={\rm log}(1/g)$.
This type of wave function is referred to as $\lambda$-function in this
paper.
In our calculations 
the second level $\lambda$-function $\psi_{\lambda}^{(2)}$
has given good results for the ground-state energy.
If we perform an extrapolation as a function of the variance, we can obtain
the correct expectation values as the QMD method.
We must, however, determine variational parameters in the multi-parameter space
by adjusting the values of the parameters to find a minimum.
The advantage of the variational procedure is that the evaluations
are stable even for large $U/t$, beyond the band width.

The correlation functions for the $4\times 4$ where $N_e=10$ and $U=4$ are
presented in Table II. The exact diagonalization results are also provided.
The correlation functions are defined as
\begin{equation}
S({\bf q})= \frac{1}{N}\sum_{ji}{\rm e}^{i{\bf q}\cdot({\bf R}_j-{\bf R}_i)}
\langle (n_{j\uparrow}-n_{j\downarrow})(n_{i\uparrow}-n_{i\downarrow})
\rangle,
\end{equation}  
\begin{equation}
C({\bf q})= \frac{1}{N}\sum_{ji}{\rm e}^{i{\bf q}\cdot({\bf R}_j-{\bf R}_i)}
(\langle n_jn_i\rangle-\langle n_j\rangle\langle n_i\rangle),
\end{equation}
\begin{equation}
s(i,j)=\langle (n_{j\uparrow}-n_{j\downarrow})(n_{i\uparrow}-n_{i\downarrow})
\rangle,
\end{equation}
\begin{equation}
c(i,j)= \langle n_jn_i\rangle-\langle n_j\rangle\langle n_i\rangle,
\end{equation}
where $n_j=n_{j\uparrow}+n_{j\downarrow}$ and ${\bf R}_j$ denotes the
position of the $j$-th site.
$\Delta_{\alpha\beta}$ is the pair correlation function,
\begin{equation}
\Delta_{\alpha\beta}(\ell)= \langle \Delta_{\alpha}^{\dag}(i+\ell)
\Delta_{\beta}(i)\rangle,
\end{equation}
where $\Delta_{\alpha}(i)$, $\alpha=x,y$, denote the annihilation operators
of the singlet electron pairs for the nearest-neighbor sites:
\begin{equation}
\Delta_{\alpha}(i)= c_{i\downarrow}c_{i+\hat{\alpha}\uparrow}
-c_{i\uparrow}c_{i+\hat{\alpha}\downarrow}.
\end{equation}
Here $\hat{\alpha}$ is a unit vector in the $\alpha(=x,y)$-direction.
The agreement in this case is good for such a small system.
The correlation functions are also dependent on the number of basis wave
functions as shown in Fig.\ref{4x4-10C}.  Since the fluctuation of the 
expectation values
is small in this case, the extrapolation can be performed in terms of 
the $1/N_{states}$.

\begin{figure}
\includegraphics[width=10.5cm]{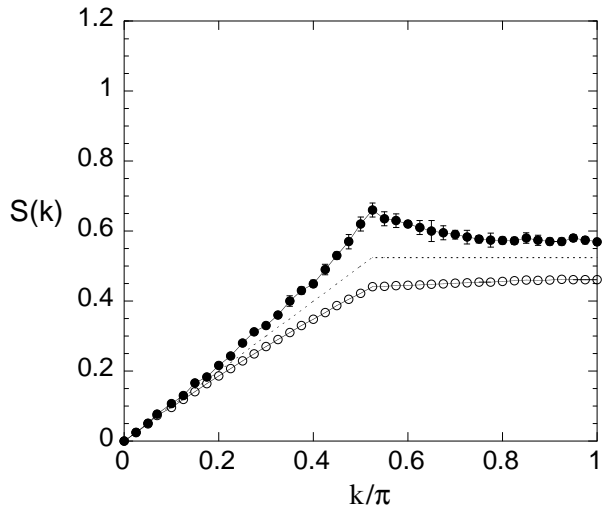}
\caption{
Spin (solid circle) and charge (open circle) correlation functions 
obtained from the QMD method for the one-dimensional
Hubbard model with $80$ sites.  The number of electrons is $66$. 
We set $U=4$ and use the
periodic boundary condition.
}
\label{1dsk}
\end{figure}

\begin{figure}
\includegraphics[width=10.5cm]{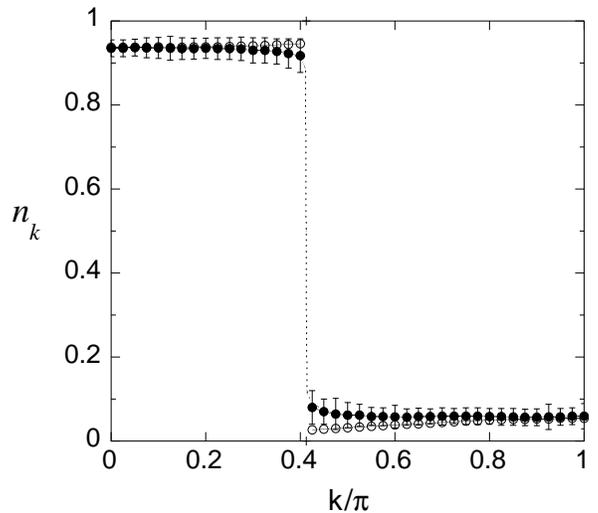}
\caption{
Momentum distribution function obtained from the QMD method for the 
one-dimensional
Hubbard model with $80$ sites for the periodic boundary condition.  
The number of electrons is $66$ and the Coulomb repulsion is $U=4$.
The dotted line is the guide given by $n_k\sim 0.5+0.4|k-k_F|^{\eta-1}$
where $\eta-1\sim 0.035$ which corresponds to $K_{\rho}\sim 0.69$ using
the formula $\eta-1=(K_{\rho}+K_{\rho}^{-1})/4-1/2$\cite{sch91}.
Open circles are the results obtained using the Gutzwiller function.
}
\label{1dnk}
\end{figure}

\begin{figure}
\includegraphics[width=9.5cm]{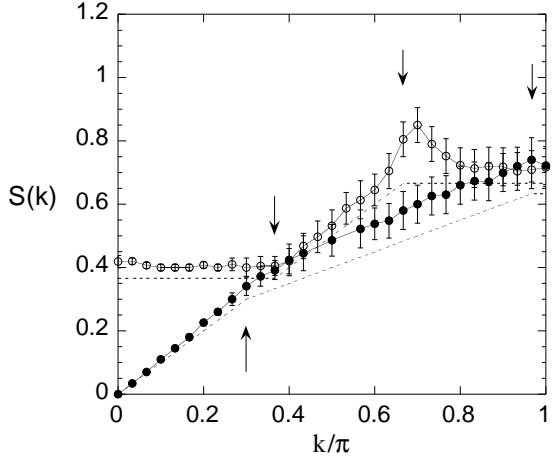}
\caption{
Spin correlation function obtained from the QMD method for the ladder
Hubbard model for $60\times 2$ sites with periodic boundary condition.
The number of electrons is $80$ and $U=4$.
The upper line is for the upper band and the lower line is for the lower band.
Singularities are at $k_{F1}-k_{F2}$, $2k_{F2}$, $k_{F1}+k_{F2}$ and
$2k_{F1}$ from left.
The dotted lines are for $U=0$.
}
\label{ldsk}
\end{figure}

\begin{figure}
\includegraphics[width=\columnwidth]{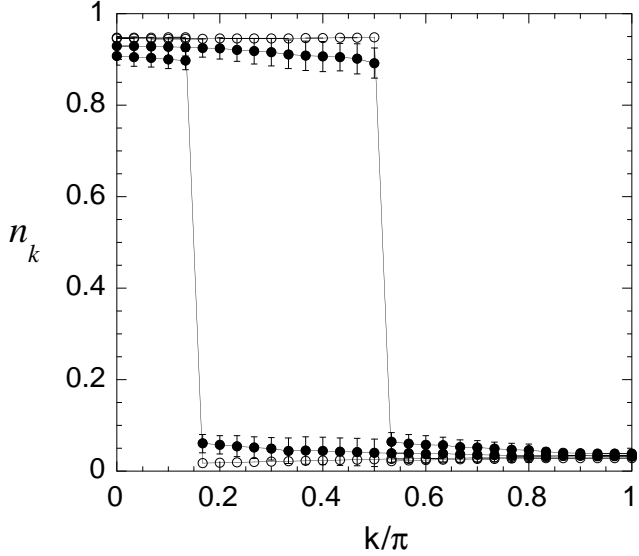}
\caption{
Momentum distribution function obtained from the QMD method for the ladder
Hubbard model for $60\times 2$ sites and periodic boundary condition.
The number of electrons is $80$ and $U=4$.
}
\label{ldnk}
\end{figure}

\begin{figure}
\includegraphics[width=10.5cm]{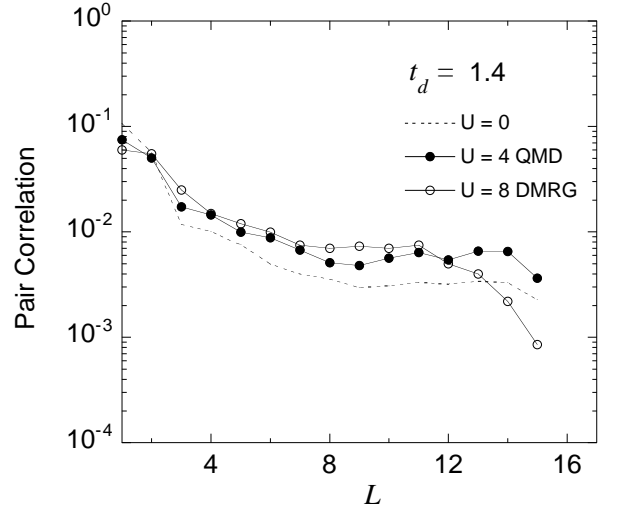}
\caption{
Pair correlation function (solid circles) obtained using the QMD method
for the ladder
Hubbard model with $16\times 2$ sites where the boundary condition
is open.
$U=4$, $t_d=1.4$ and the electron filling is 0.875.
The dashed line is the pair correlation function for $U=0$.
The open circles are the DMRG results from Ref.\cite{noa97}.
}
\label{ldpair}
\end{figure}

\begin{figure}
\includegraphics[width=\columnwidth]{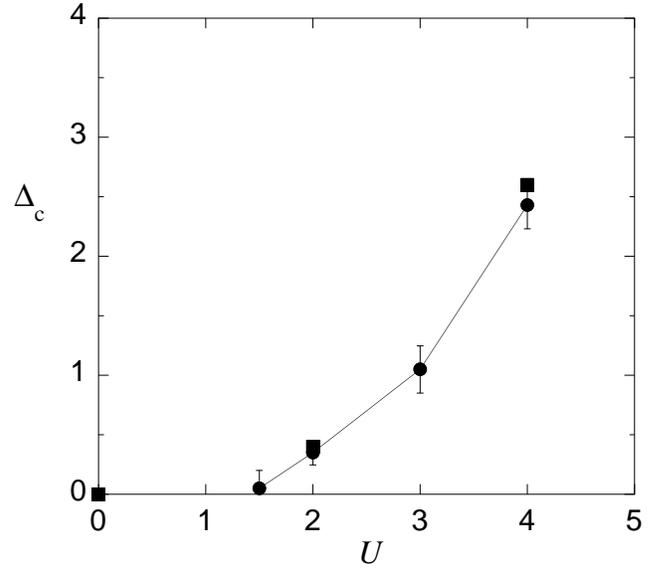}
\caption{
Charge gap as a function of $U$ for $t_d=1$ (circles).
The DMRG results (squares) are provided for comparison\cite{dau00}.
}
\label{ldDc}
\end{figure}

\begin{figure}
\includegraphics[width=\columnwidth]{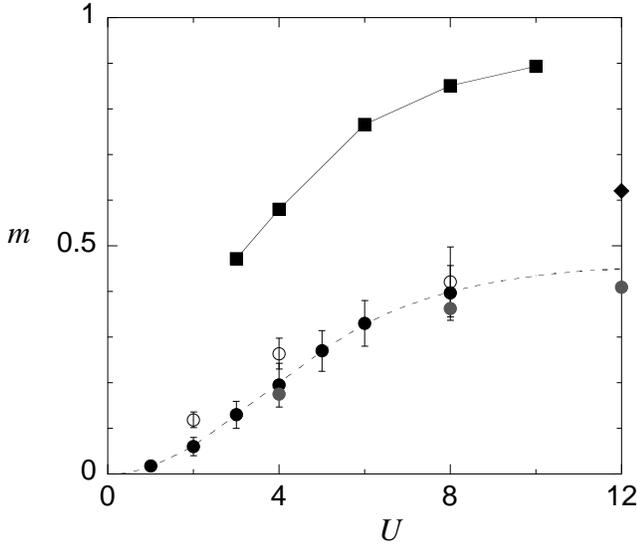}
\caption{
Magnetization as a function of $U$ for the half-filled Hubbard model
after extrapolation at the limit of large $N$.
Solid circles are the QMD results, and open circles are results obtained 
from the QMC method\cite{hir85}.  The squares are the Gutzwiller-VMC 
results\cite{yok87} and gray solid
circles show the 3rd $\lambda$-function ($\psi_{\lambda}^{(3)}$) VMC results 
carried out on the $8\times 8$ lattice\cite{yan98}.
The diamond symbol is the value from the two-dimensional Heisenberg model
where $m=0.615$\cite{rie89,cal98}.
}
\label{m-U}
\end{figure}

\begin{figure}
\includegraphics[width=10.5cm]{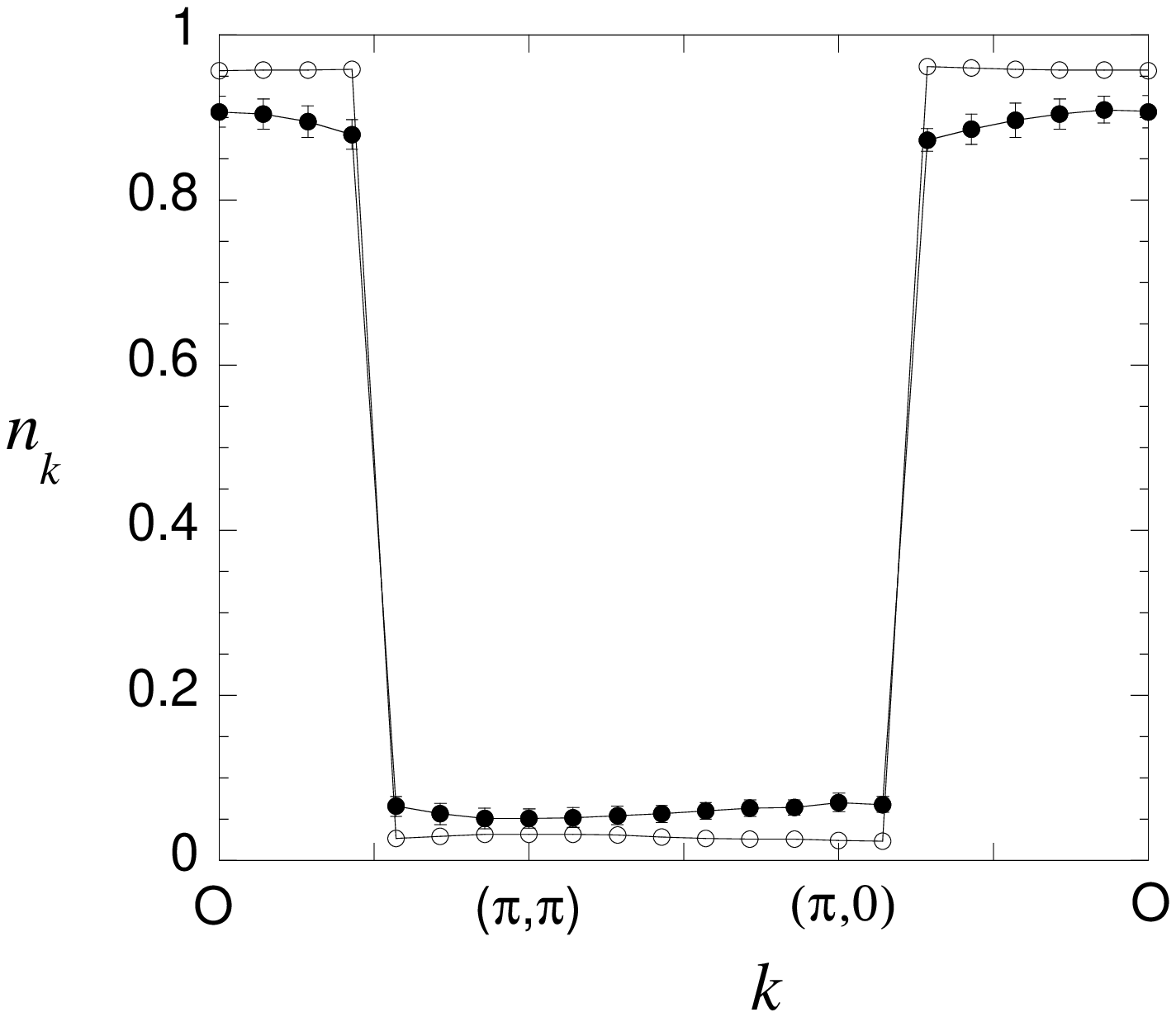}
\caption{
Momentum distribution function for the $14\times 14$ lattice.
Parameters are $U=4$ and $N_e=146$.  The boundary conditions are periodic
in both directions.
The results for the Gutzwiller function (open circle) are also provided.
}
\label{2Dnk}
\end{figure}

\subsection{1D and Ladder Hubbard models}

In this subsection we show the results for the one-dimensional (1D) Hubbard
model and ladder Hubbard model.  
The ground state of the 1D Hubbard model is no longer Fermi liquid for $U>0$.
The ground state is insulating at half-filling and metallic for less than
half-filling.
The Fig. \ref{1dsk} is the spin and charge correlation 
functions, $S(k)$ and $C(k)$,
as a function of the wave number, for the 1D Hubbard model where $N=80$.  
The $2k_F$ singularity can be clearly identified where the dotted line is
for $U=0$. 
The spin correlation is enhanced and the charge correlation function is
suppressed slightly  because of the Coulomb interaction.
The momentum distribution function $n(k)$,
\begin{equation}
n(k)= \frac{1}{2}\sum_{\sigma}\langle c_{k\sigma}^{\dag}c_{k\sigma}\rangle,
\end{equation}
is presented in Fig.\ref{1dnk} for the electron filling $n=0.825$.
Here $c_{k\sigma}$ is the Fourier transform of $c_{j\sigma}$.
$n(k)$ in the metallic phase exhibits a singular behavior near the wave number $k_F$.
The singularity close to $k_F$ is consistent with the property of the Luttinger
liquid\cite{sch91,kaw90}, although
it is difficult to analyze the singularity in more detail using the
Monte Carlo method.
The Gutzwiller function gives the unphysical result that $n(k)$ increases
as $k$ approaches $k_F$ from above the Fermi surface.

In the ladder Hubbard model,
\begin{eqnarray}
H_{ladder}&=& 
-t\sum_{\ell=1,2}\sum_{j\sigma}(c^{\dag}_{\ell j\sigma}c_{\ell j+1,\sigma}
+{\rm h.c.})\nonumber\\
&-&t_d\sum_{j\sigma}(c^{\dag}_{1j\sigma}c_{2j\sigma}+{\rm h.c.})\nonumber\\
&+&U\sum_{\ell=1,2}\sum_j c^{\dag}_{\ell j\uparrow}c_{\ell j\uparrow}
c^{\dag}_{\ell j\downarrow}c_{\ell j\downarrow},
\end{eqnarray}
where $t(t_d)$ is the intrachain (interchain) transfer energy.
The ladder Hubbard model exhibits a spin gap at half-filling, and the
charge gap is also possibly opened for large $U>0$ at half-filling.
The existence of superconducting phase has been suggested for the 
Hubbard ladder using the DMRG method\cite{noa97} and the VMC method\cite{koi99}.

The spin correlation function $S({\bf k})$ for the Hubbard ladder is presented in 
Fig.\ref{ldsk},
where $U=4$ and $t_d=1$.
$S({\bf k})$ is defined as
\begin{equation}
S({\bf k})= \frac{1}{N}\sum_{i\ell,j\ell'}{\rm e}^{i{\bf k}\cdot({\bf R}_{i\ell}
-{\bf R}_{j\ell'})}\langle(n_{\ell i\uparrow}-n_{\ell i\downarrow})
(n_{\ell' j\uparrow}-n_{\ell' j\downarrow})\rangle,
\end{equation}
where ${\bf R}_{i\ell}$ denotes the site $(i,\ell)$ ($\ell$=1,2). 
We use the convention that ${\bf k}=(k,k_y)$ where $k_y=0$ and $\pi$
indicate the lower band and upper band, respectively.
There are four singularities at $2k_{F1}$, $2k_{F2}$, $k_{F1}-k_{F2}$, 
and $k_{F1}+k_{F2}$ for the Hubbard ladder, where $k_{F1}$ and 
$k_{F2}$ are the 
Fermi wave numbers
of the lower and upper band, respectively.
They can be clearly identified as indicated by arrows in Fig.\ref{ldsk}.

The momentum distribution in Fig.\ref{ldnk} 
\begin{equation}
n({\bf k})= \frac{1}{2N}\sum_{\sigma}
\sum_{i\ell,j\ell'}{\rm e}^{i{\bf k}\cdot({\bf R}_{i\ell}-{\bf R}_{j\ell'})}
\langle c^{\dag}_{\ell i\sigma}c_{\ell' j\sigma}\rangle,
\end{equation}
exhibits singularities at
$k_{F1}$ and $k_{F2}$ where the results obtained from the Gutzwiller function
are also shown for comparison.
Here we used the same notation for ${\bf k}$ and ${\bf R}_{i\ell}$.
The unphysical property of $n({\bf k})$ near the Fermi wave numbers for the
Gutzwiller function are remedied in the QMD method.

The pair correlation function, $\Delta_{yy}(\ell)$ versus $\ell$ was also
evaluated to compare with the density matrix renormalization group (DMRG) 
method.  $\Delta_{yy}(\ell)$ is defined as
\begin{equation}
\Delta_{yy}(\ell)= \langle \Delta_y^{\dag}(i+\ell)\Delta_y(i)\rangle
\end{equation}
for 
\begin{equation}
\Delta_y(i)= c_{1i\downarrow}c_{2i\uparrow}-c_{1i\uparrow}c_{2i\downarrow}.
\end{equation}
$\Delta_{yy}(\ell)$ is the correlation function for the singlet pair on 
the rung.
The results for $\Delta_{yy}(\ell)$ are given in 
Fig.\ref{ldpair} on the $16\times 2$ lattice for the open boundary condition,
where the pair correlation 
functions $\Delta_{yy}(\ell)$ were averaged over several pairs, for a 
distance $\ell$.
The values $U=4$ and $t_d=1.4$ are predefined, and the 
electron filling was $n=0.875$.
The result obtained using the DMRG method is also
provided for $U=8$\cite{noa97} for comparison.
Since a large value of $U$, such as $U=8$, is not easily accessed using
the QMD method, we have presented the results for $U=4$.
The enhancement of the pair correlation function over the non-interacting case 
is clear and is consistent with the DMRG method.

It has been expected that the charge gap opens up as $U$ turns on at half-filling
for the Hubbard ladder model.
In Fig.\ref{ldDc} the charge gap at half-filling is shown as a function of $U$.
The charge gap is defined as
\begin{equation}
\Delta_c= E(N_e+2)+E(N_e-2)-2E(N_e),
\end{equation}
where $E(N_e)$ is the ground state energy for the $N_e$ electrons.
The charge gap in Fig.\ref{ldDc} was estimated using the extrapolation to 
the infinite
system from the data for the $20\times 2$, $30\times 2$, and $40\times 2$
systems.
The data are consistent with the DMRG method and suggest the exponentially small
charge gap for small $U$ or the existence of the 
critical value $U_c$ in the range
of $0\leq U_c <1.5$, below which the charge gap vanishes.

\subsection{2D Hubbard model}

The two-dimensional Hubbard model was also investigated in this study.
The results are presented in the following discussion.
An important issue is the antiferromagnetism at half-filling.
The ground state is antiferromagnetic for $U>0$ because of the nesting
due to the commensurate vector $Q=(\pi,\pi)$.
The Gutzwiller function predicts that the magnetization
\begin{equation}
m=\left|\frac{1}{N}\sum_j(n_{j\uparrow}-n_{j\downarrow})
{\rm e}^{iQ\cdot R_j}\right|
\end{equation}
increases rapidly as $U$ increases and approaches $m=1$ for large $U$.
In Fig.\ref{m-U} the QMD results are presented for $m$ as a function of $U$.
The previous results obtained using the QMC method are plotted as open circles.
The gray circles are for the $\lambda$-function VMC method and squares
are the Gutzwiller VMC data.  Clearly, the magnetization is reduced considerably
because of the fluctuations, and is smaller than the Gutzwiller VMC method by 
about 50 percent.

The Fig. \ref{2Dnk} is the momentum distribution function $n({\bf k})$,
\begin{equation}
n({\bf k})= \frac{1}{2}\sum_{\sigma}\langle c_{{\bf k}\sigma}^{\dag}
c_{{\bf k}\sigma}\rangle,
\end{equation}
where the results for the Gutzwiller VMC and
the QMD are indicated.
The Gutzwiller function gives the results that $n(k)$ increases as $k$
approaches $k_F$ from above the Fermi surface.
This is clearly unphysical.
This flaw of the Gutzwiller function near the Fermi surface is not
observed for the QMD result.

\begin{table}
\caption{Ground state energy per site from the Hubbard model. 
The boundary conditions are periodic in both directions.  
The current results are presented under the column labeled QMD.
The constrained
path Monte Carlo (CPMC) and Path integral renormalization group (PIRG) results
are from Refs.\cite{zha97} and \cite{kas01}, respectively.  
The column VMC is the results obtained for the optimized variational
wave function $\psi_{\lambda}^{(2)}$ except for $6\times 2$ for which
$\psi_{\lambda}^{(1)}$ is employed.
The QMC results are from Ref.\cite{fur92}.
Exact results are obtained using 
diagonalization\cite{par89}.}
\begin{tabular}{ccccccccc}
\colrule
Size & $N_e$ & $U$ &  QMD & VMC & CPMC & PIRG & QMC & Exact \\
\colrule
$4\times 4$ & 10 & 4  & -1.2237   & -1.221(1) & -1.2238 &  &    & -1.2238 \\
$4\times 4$ & 14 & 4  & -0.9836   & -0.977(1) & -0.9831 &  & & -0.9840 \\
$4\times 4$ & 14 & 8  & -0.732(2) & -0.727(1) & -0.7281 &  & & -0.7418 \\
$4\times 4$ & 14 & 10 & -0.656(2) & -0.650(1) &         &  & & -0.6754 \\
$4\times 4$ & 14 & 12 & -0.610(4) & -0.607(2) & -0.606  &  & & -0.6282 \\
$6\times 2$ & 10 & 2  & -1.058(1) & -1.040(1) &         &  & & -1.05807\\
$6\times 2$ & 10 & 4  & -0.873(1) & -0.846(1) &         &  & & -0.8767 \\
$6\times 6$ & 34 & 4  & -0.921(1) & -0.910(2) &         & -0.920  & -0.925 & \\
$6\times 6$ & 36 & 4  & -0.859(2) & -0.844(2) &         & -0.8589 & -0.8608 & \\
\colrule
\end{tabular}
\end{table}

\section{Summary}
We have presented a Quantum Monte Carlo diagonalization method for a
many-fermion system.
We employ the Hubbard-Stratonovich transformation to decompose the
interaction term as in the standard QMC method.  We use this in an expansion
of the true ground-state wave function.
We have considered the truncated space of the basis functions $\{\phi_m\}$
and diagonalize the Hamiltonian in this subspace.
We can optimize the wave function by enlarging the subspace.
The simplest way is to increase the number of basis functions by
randomly generating auxiliary fields $\{s_i\}$.
The wave function can be further improved by multiplying each $\phi_m$ by
 $B_{\ell}^{\sigma}$.
Although the matrix $B_{\ell}^{\sigma}$ in Eq.(\ref{bmat}) generates $2^N$ 
new basis
functions, we must select some states from them to keep the number of
basis functions small.
Within the subspace with the fixed number of basis functions,
an extension of $1/2^N$-method to $k/2^N$ method ($k=1,2,\cdots$) is also 
possible.

We have proposed a genetic-algorithm based method to generate the basis wave
functions.  The genetic algorithm is widely used in solving problems
to find the optimized solution in the space of large configuration numbers.
We make new basis functions from the functions with large weighting factors
$\left|c_n\right|^2$.  New functions produced in this way are expected to have
large weighting factors.
If the localization quotient $Q_{loc}$ in Eq.(\ref{qloc}) is not small,
we can iterate the Monte Carlo steps without using the $1/2^N$-method.

We have computed the energy and correlation functions for small lattices
to compare with published data.
The results obtained in this study are consistent with the published data.
In the case of the open shell structures, evaluations are difficult in
general and the convergence is not monotonic. 
In this case the subspace of the basis functions must be large to obtain
the expectation values from the
extrapolation procedure.

As for the extrapolation, the expectation value $\langle Q\rangle$ may 
approach $Q_{exact}$
in a non-linear way,
\begin{equation}
\langle Q\rangle-Q_{exact}\propto (N_{states})^{-\theta}
\end{equation}
for some exponent $\theta$.
We must evaluate $\theta$ to obtain $Q_{exact}$, from an extrapolation
in terms of the $N_{states}^{-1}$.  
We may be able to use a derivative method where $\theta$ is determined
so that the derivative $d\langle Q\rangle/dN_{states}$ approaches 0
as $N_{states}$ increases.
In this paper we adopted the recently proposed energy-variance
method\cite{sor01,kas01}.
For the energy and local quantities, we can expect 
$\langle Q\rangle-Q_{exact}\propto v$ for the variance $v$.
It is expected that the long-range correlations are not trivial to calculate
since the orthogonality $\langle\psi_i Q\psi_g\rangle\approx 0$ should hold
for the ground state $\psi_g$ and excited states $\psi_i$.

\section{Acknowledgments}
  We thank J. Kondo, K. Yamaji and S. Koikegami for helpful discussions.

\begin{table}
\caption{Correlation functions for the $4\times 4$ Hubbard model with periodic 
boundary conditions.
Parameters are $N_e=10$ and $U=4$.
VMC indicates the variational Monte Carlo results obtained by 
$\psi_{\lambda}^{(2)}$.  
CPMC indicates the constrained path Monte Carlo results.}
\begin{tabular}{ccccc}
\colrule
Correlation function  & QMD  &   VMC     &  CPMC  &  Exact \\
\colrule
$S(\pi,\pi)$     &  0.730(1)  &  0.729(2) & 0.729  &  0.7327  \\
$C(\pi,\pi)$     &  0.508(1)  &  0.519(2) & 0.508  &  0.5064  \\
$\Delta_{yy}(1)$ &  0.077(1)  &  0.076(1) &        &  0.07685  \\
$\Delta_{yy}(2)$ &  0.006(1)  &  0.006(1) &        &  0.00624 \\
$\Delta_{xy}(0)$ &  0.124(1)  &  0.120(2) &        &  0.1221   \\
$\Delta_{xy}(1)$ & -0.015(1)  & -0.015(1) &        & -0.0141  \\
$s(0,0)$         &  0.529(1)  &           &        &  0.5331  \\
$s(1,0)$         & -0.091(1)  &           &        & -0.0911  \\
$c(0,0)$         &  0.329(1)  &           &        &  0.3263  \\
$c(1,0)$         & -0.0536(1) &           &        & -0.05394 \\
\colrule
\end{tabular}
\end{table}

\end{document}